\documentclass[aps,superscriptaddress,prl]{revtex4}

\usepackage[english]{babel}
\usepackage[utf8]{inputenc}
\usepackage{amsmath,amssymb}
\usepackage{graphicx}
\usepackage{epsfig}
\usepackage[colorinlistoftodos]{todonotes}
\usepackage{psfrag}
\usepackage[colorlinks=true,breaklinks=true]{hyperref}
\DeclareMathOperator{\Grad}{Grad}

\begin{document}
\title{Modelling afferent nerve responses to bladder filling*}

\author{Maryam Argungu}
\affiliation{Department of Bioengineering, Imperial College London,  London, UK}

\author{Saziye Bayram}
\affiliation{Department of Mathematics, SUNY Buffalo State, Buffalo NY, USA}

\author{Bindi Brook}
\affiliation{School of Mathematical Sciences,  University of
  Nottingham, Nottingham, UK}

\author{Buddhapriya Chakrabarti}
\affiliation{Department of Mathematical Sciences, Durham University,
  Durham, UK}

\author{Richard H Clayton}
\affiliation{Department of Computer Science, University of Sheffield, Sheffield, UK}

\author{Donna M Daly}
\affiliation{Department of Biomedical Science, University of Sheffield,
  Sheffield, UK}

\author{Rosemary J Dyson}
\author{Craig Holloway}
\affiliation{School of Mathematics, University of Birmingham, Birmingham, UK}

\author{Varun Manhas}
\affiliation{Biomedical Engineering, University of Li\'ege \& KU Leuven, Leuven, Belgium}

\author{Shailesh Naire}
\affiliation{School of Computing and Mathematics, Keele University,
  Keele, UK}

\author{Tom Shearer}
\affiliation{School of Mathematics University of Manchester, Manchester, UK}

\author{Radostin Simitev}
\affiliation{School of Mathematics and Statistics, University of Glasgow, Glasgow, UK}

\date{\today}

{\small * This report summarises the outcomes from a problem posed at
  the UK MMSG NC3R's \& POEMS Study Group meeting, 8--12 Sept
  2014, Cambridge}\vspace{1cm}

\begin{abstract}
A sensation of fullness in the bladder is a regular experience, yet the mechanisms that act to generate this sensation remain poorly understood. This is an important issue because of the clinical problems that can result when this system does not function properly. The aim of the study group activity was to develop mathematical models that describe the mechanics of bladder filling, and how stretch modulates the firing rate of afferent nerves. Several models were developed, which were qualitatively consistent with experimental data obtained from a mouse model.
\end{abstract}

\maketitle

\section{Introduction}

The urinary bladder acts to collect and store urine at low pressure, with controlled emptying when the bladder becomes full \cite{Andersson2004}. Anatomically it is a hollow muscular organ, with 4 concentric layers. On the innermost surface is a thin epithelial layer (termed the urothelium) surrounded by a laminar propria (or suburothelium) and a thick smooth muscle layer (the detrusor) which is covered by an elastin rich serosa. The thick detrusor muscle wall is formed by smooth muscle cells. Afferent nerves within the bladder respond to chemical and mechanical stimuli, and act to generate a sensation of fulness \cite{deGroat2009}. 

Urine is produced by the kidneys and delivered to the bladder via two thin muscular tubes called ureters, where it is stored until emptying. As the bladder fills, it deforms to accommodate increases in urine volume without a dramatic rise in pressure. On emptying, the detrusor muscle contracts and urine is eliminated from the bladder via the urethra. In order to achieve continence during bladder filling and storage and produce efficient and effective bladder emptying, there is accurate co-ordination between opening and closing of sphincters located in the urethra and contraction of the detrusor smooth muscle. The process of micturition has two phases, a storage/filling phase and a voiding phase. These phases are regulated by a complex integration of somatic and autonomic efferent and afferent mechanisms that coordinate the activity of the bladder and urethra. The transition to synchronous bladder contraction and urethral outlet relaxation respectively, and vice versa, is co-ordinated and under voluntary control of the central nervous system \cite{deGroat1993,deGroat2009}.

Alterations in this normal cycle of filling and emptying, or changes to the function of any of the governing components can give rise to a number of clinical conditions such as overactive bladder syndrome (OAB) and urinary incontinece (UI). Until recently, research focussed on the mechanisms which drive bladder contractility and initate smooth muscle function. However it has now become clear that the peripheral sensory nerves which detect bladder filling and trigger the micturition cycle may drive the symptoms of these disorders and could even be attractive drug targets. This has led to studies looking at the function of these nerves in animals. However a detailed and integrative understanding of the interactions between the different signalling mechanisms remains elusive.

Afferent nerves innervate, and have terminal endings in all of the layers of the bladder and form a dense neural plexus between the urothelium and laminar propria. On filling, intravesical pressure increases with increased volume, triggering activation of mechanosensitive nerve fibres (those responsive to stretch of the tissue or distortion of the nerve terminal endings). These nerves are classified according to their innervation pattern, stimulus-response profile and/or threshold for activation. Experimental studies suggest that there are more than 5 distinct sub-populations of afferent nerve units innervating the bladder. These include both stretch sensitive and stretch insensitive afferents (i.e. chemosensitive afferents that only respond to chemical mediators) \cite{Zagorodnyuk2006}. 

\begin{figure}
\includegraphics[height=\textwidth,angle=-90]{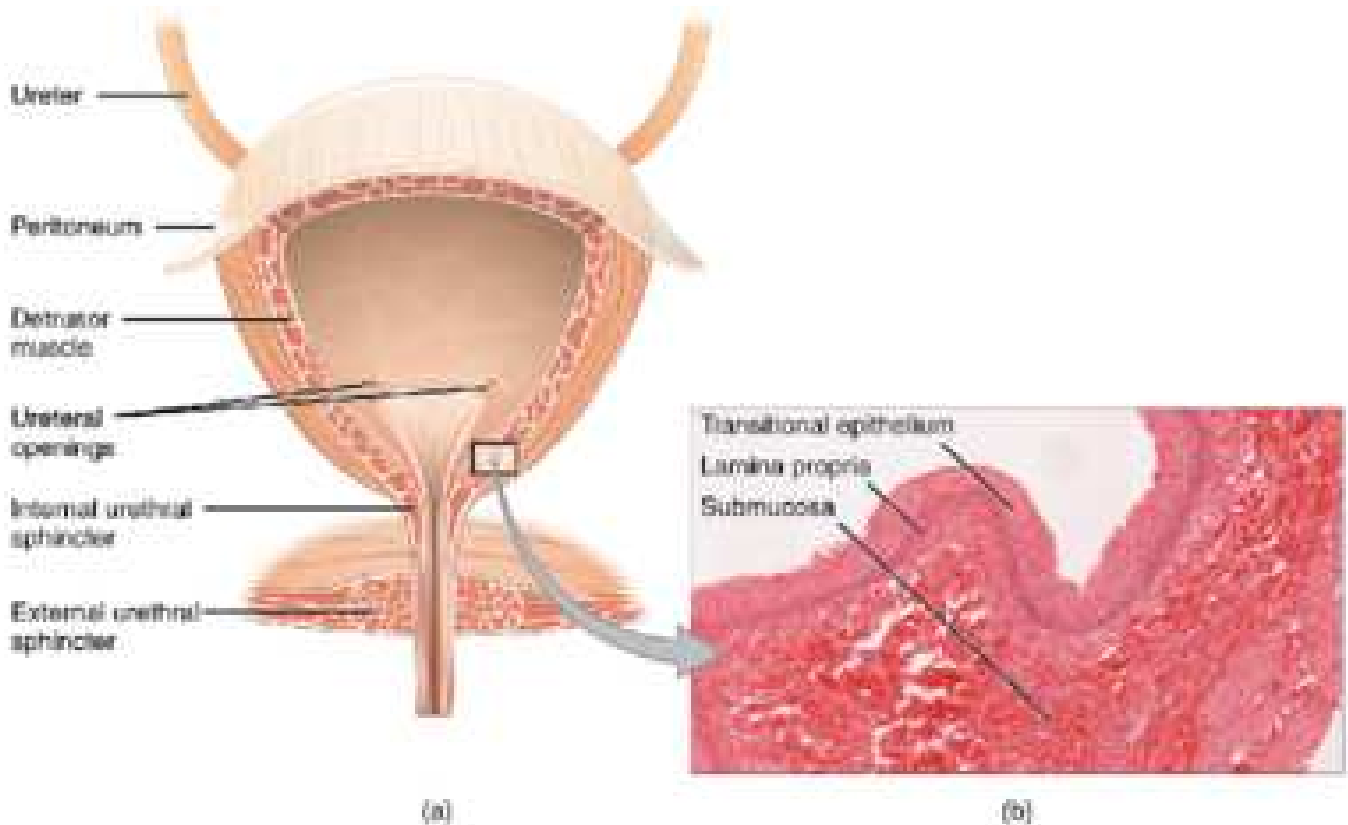}
	
\caption{Anatomy of the bladder, showing (a) gross anatomy of the huma female bladder, and (b) a cross section of the bladder wall, showing different layers of tissue. Reproduced with	permission from Anatomy \& Physiology, Connexions Web site. \url{http://cnx.org/content/col11496/1.6/} }
	
\label{bladderWall}
\end{figure}

The urothelium is a 5-7 cell thick layer of epithelium which lines the luminal surface of the bladder (Figure~\ref{bladderWall}). Its principal role is to act as a barrier protecting the underlying tissues from the noxious acidic urine. However studies have also shown that in addition to its barrier function the urothelium is also able to detect bladder filling and modulate or tune afferent signals via the release of both excitatory and inhibitory chemical mediators. These mediators act on nearby afferent nerve terminals to modulate or activate nerve activity (via both mechanosenstive and  chemosentive afferents). The urothelium also plays a key role in accomodation of the bladder to increasing volumes. It consists of large apical cells termed umbrella cells, intermediate cells and small basal cells. During filling, preformed membrane segments are inserted into the umbrella cell membrane to increase surface area of these cells increasing luminal volume. As yet there are no reliable models (experimental or otherwise) which allow us to predict how manipulations of the bladder, such as a changes in distensibility associated with age, alter afferent nerve function. Moreover measuring the effect of stimulus on a single nerve fibre is incredibly difficult and requires the use of large numbers of animals. 

\section{Description of the problem}


The aim of this project was to begin to develop mathematical models, which simulate bladder filling and predict what effect this has on the distinct subpopulations of mechanosensitive afferent nerves innervating all 4 of the concentric layers of the bladder. Moreover the models should also take into account the contribution of urothelium distension and signalling to this overall afferent output.

Mathematical models which emcompass these considerations would provide insight into how the distinct subpopulations of afferent nerve fibres that terminate in different layers of the bladder wall respond to  bladder filling, and how these signals are combined to produce overall sensory input to the micturition circuitry. A fully integrated model would also allow us to predict how structural changes such as those associated with ageing (i.e. collagen deposition/loss, increased/decreased muscle loss) influence mechanical properties of the bladder, and in turn the sensory response to bladder filling. A further benefit of this approach is the potential to examine the contribution of urothelial signalling to these pathways. An important role of modelling is to frame hypotheses and to identify and refine novel future experiments.

During the study group week, we worked on models of two aspects of bladder structure and function: (i) whole bladder filling and emptying and (ii) mechanosensitive nerve activation as a result of filling. A number of different approaches were used to investigate these aspects and are detailed below.






\section{Mechanical Model of Whole Bladder Filling}

As a first stage, a simple lumped parameter model was developed to understand basic flow and structure mechanics in the bladder. The simple model also accounted for afferent nerve firing associated with bladder filling. 

We begin by summarising the model variables, parameters, and assumptions and then provide the details of the model.


\vspace{0.1in}
Model Variables:

$V_b(t)$ = Volume inside the bladder

$Q_{in}(t)$ = Flux into the bladder (from two ureters collectively)

$Q_s(t)$ = Flux out from the bladder through sphincter

$P_b(t)$ = Pressure inside the bladder

$N(t)$ = Firing rate of afferent nerves

\vspace{0.1in}
Model Parameters:

$R_s$ = Resistance at the sphincter

$P_s$ = Pressure at the sphincter

$P_e$ = Pressure outside the bladder

$C$ = Bladder wall compliance

\vspace{0.1in}
Model assumptions at $t=0$:

$Q_s$=0

$P_e$=0

$V_b$=$V_{b_{0}}$

\vspace{0.1in}



The change in bladder volume, $dV_b/dt$, can be modelled as a one dimensional first order differential equation. In this case, we ignore the shape of the bladder and consider the volume and pressure balance in the system.

\begin{equation} 
\frac{\mathrm{d}V_b(t)}{\mathrm{d}t} =Q_{in}(t) - Q_s(t),
\end{equation}

\noindent where $Q_{in}$ and $Q_s$ are the fluxes in and out of the bladder via two ureters and sphincter, respectively. The  pressure in the bladder, $P_b$, is related to the compliance of the bladder $dV_b/dP_b$ and may be related to the rate of change of volume $dV_b/dt$
\begin{equation}
P_b - P_e = C(V_b) + B \frac{\mathrm{d}V_b}{\mathrm{d}t},
\end{equation}
where $P_e$ is the external pressure of the bladder, and B is a constant.

\vspace{0.1in}
\noindent Emptying of the bladder via the sphincter is given by 
\begin{equation}
P_b = P_s + R_sQ_s .
\end{equation}

\noindent During filling, ($Q_s$ =0), and we assume that  $P_e$ =0 so that

\begin{equation}
V_b(t) = V_{b_{0}} + Q_{in}t
\end{equation}
\begin{equation}
P_b =C(V_{b_{0}} +Q_{in}t) + BQ_{in} .
\end{equation}

\noindent We define the firing rate of the nerves in response to bladder pressure and volume, $N(t)$, as

\begin{equation}
N(t) = k_1n_e -k_2n_i + N_0 \left (1- e^{\frac{-P_b(V_b -V_{CN})}{K T}} \right),
\end{equation}

\noindent where $n_i$ and $n_e$ are inhibitory and excitatory effects respectively. These two effects in turn depended on bladder  pressure and volume, and were modelled as 
\begin{equation}
 n_i = \left\{ 
  \begin{array}{l l}
    0 & \quad  V_b < V_{c_{i}}\\
     n_{i_{0}} \left (1- e^{\frac{-P_b(V_b -V_{c_{i}})}{K T}} \right)& \quad V_b > V_{c_{i}},
  \end{array} \right.
 \end{equation}
 
\noindent and 
\begin{equation}
 n_e = \left\{ 
  \begin{array}{l l}
    0 & \quad  V_b < V_{c_{e}}\\
     n_{e_{0}} \left (1- e^{\frac{-P_b(V_b -V_{c_{e}})}{KT}} \right)& \quad V_b > V_{c_{e}},
  \end{array} \right.
 \end{equation}

where $V_{c_{i}}$ and $V_{c_{e}}$ are threshold volumes at which the inhibitory and excitatory effects switch on.

Although this simplified mechanical model was not pursued further, it  allowed us to discuss and establish a common understanding of the underlying physiology and mechanisms so as to develop the more detailed models described in the following section.

\section{Modelling mechano-sensitive response in afferent nerve}


\subsection{Description of mathematical model}

\subsubsection{Overall approach}

\begin{figure}
\hspace*{7mm}
\includegraphics[width=0.7\textwidth]{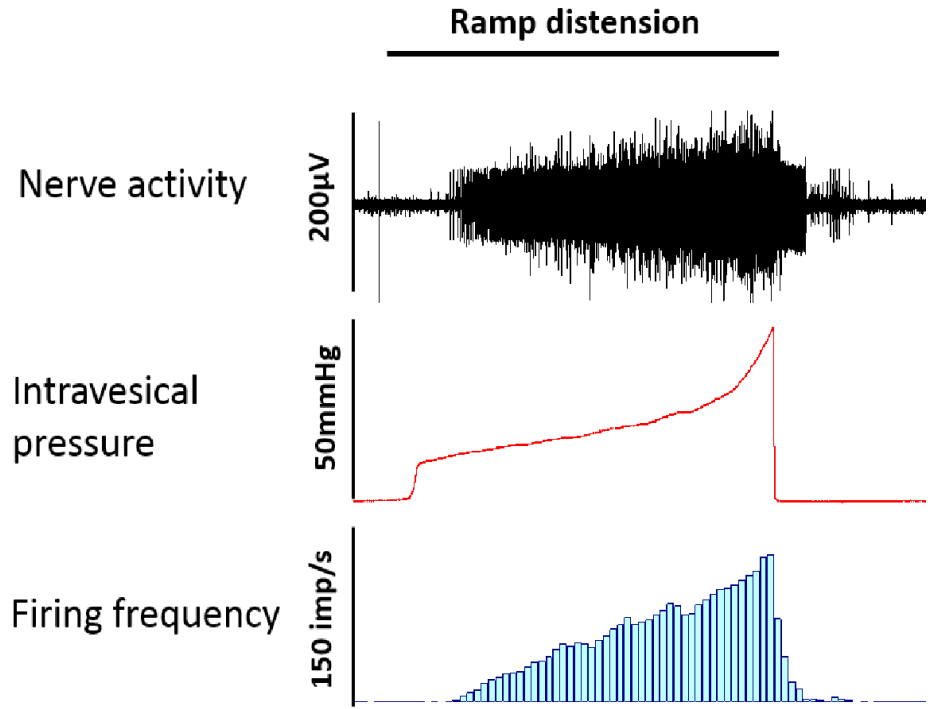}
\caption{Experimental data from mouse bladder \cite{Daly2014}, showing nerve activity (top), bladder pressure(middle), and firing frequency (bottom).}
\label{NerveFiring1}
\end{figure}

\begin{figure}
\vspace*{20mm}\hspace*{7mm}
\includegraphics[width=0.2\textwidth]{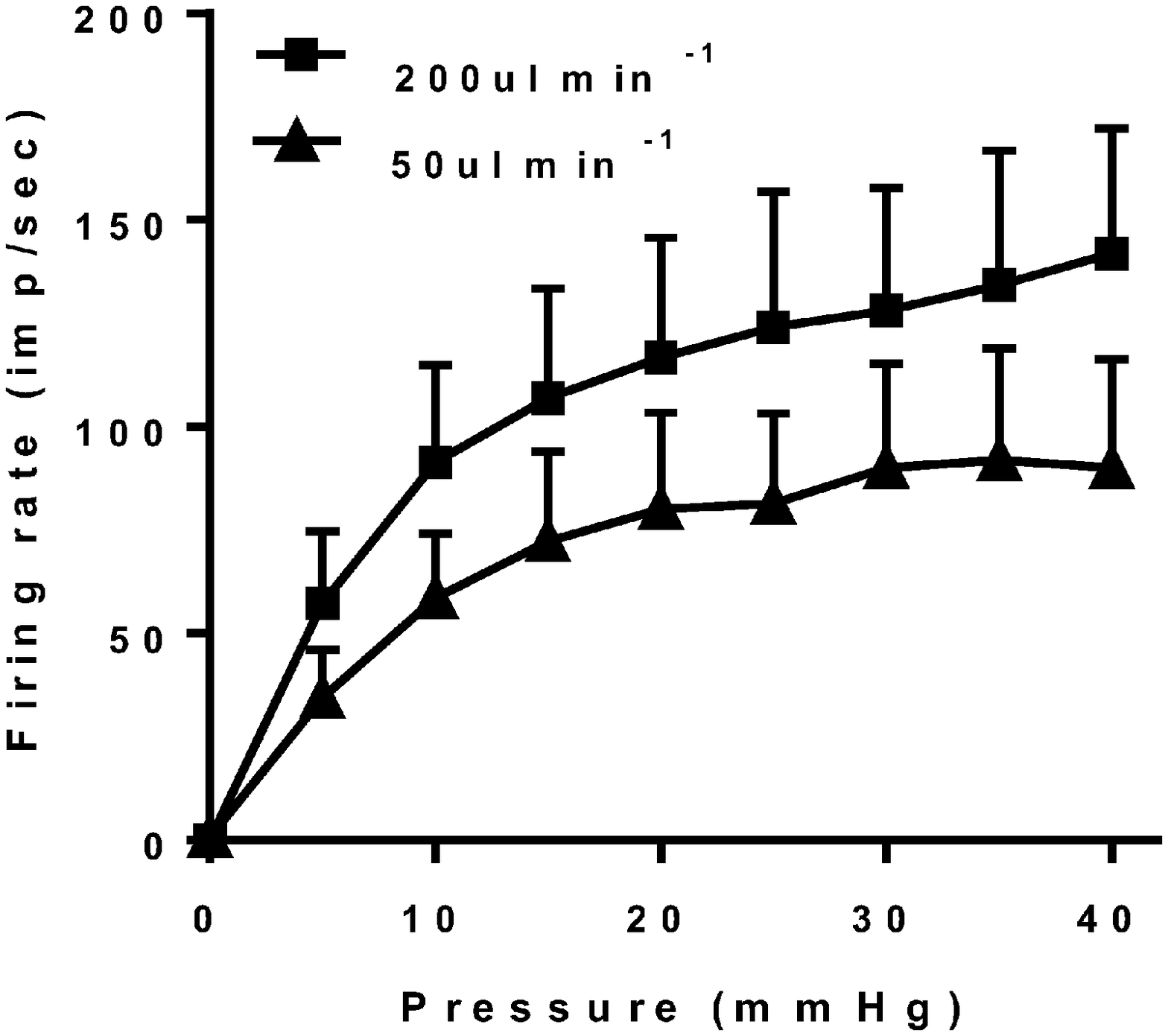}
\caption{Experimental data from mouse bladder \cite{Daly2014}, showing nerve activity for two different filling rates. }
\label{NerveFiring2}
\end{figure}

A key aspect of the problem was to develop models that describe the behaviour of stretch sensitive mechanoreceptors. Experimental data from the mouse bladder shows that the firing rate of these neurons increases as the bladder is filled. This is ilustrated in figures \ref{NerveFiring1} and \ref{NerveFiring2}. One biophysical mechanism that is believed to underlie this response is stretch activated ion channels, where the ion channel conductance increases as the cell membrane is stretched. The aim of this work was therefore to \textbf{include a mechano-sensitive ion channel} current in a model of the neural action potential so that \textbf{a bifurcation from excitable to oscillatory dynamics occurs} when the neuronal membrane is stretched.


\subsubsection{Models of the neural action potential}
Typical models of the neural single-cell action potential take the
form of a Tikhonov fast-slow system where $v$ and $y$ are state variables (for an excitable cell $v$ represents membrane voltage and not volume) and $\epsilon$ a constant
\begin{align}
\label{eq1}
&\frac{d v}{d t} = f (v,y),\\
&\frac{d y}{d t} =\epsilon g (v,y), \qquad \epsilon\ll 1. \nonumber
\end{align}

The nullcline $\dot{v}=0$ has two stable branches separated by an unstable
branch which allows for
\begin{itemize}
\item a stable equilibrium if the
  intersection of $\dot{v}=0$ and $\dot{y}=0$ is on a stable branch \textbf{(excitable                                          case)}, OR 
\item a stable
limit cycle if the intersection of $\dot{v}=0$ and $\dot{y}=0$ is on the
unstable branch \textbf{(oscillatory case)}.
\end{itemize}


\subsubsection{Phase portrait deformation}
The bifurcation from the excitable to the oscillatory case can be
achieved by including in \eqref{eq1} a mechano-sensitive current
$I_M(\lambda,v)$ that will act to deform its phase portrait
\begin{align}
\label{eq2}
&\frac{d v}{d t} = f (v,y) + {I_M(\lambda,v)},\\
&\frac{d y}{d t} =\epsilon g (v,y), \nonumber
\end{align}
where $\lambda$ is a suitable ``mechanical'' parameter.

\subsection{Model of the mechano-sensitive current}

\begin{figure}
\includegraphics[width=\textwidth]{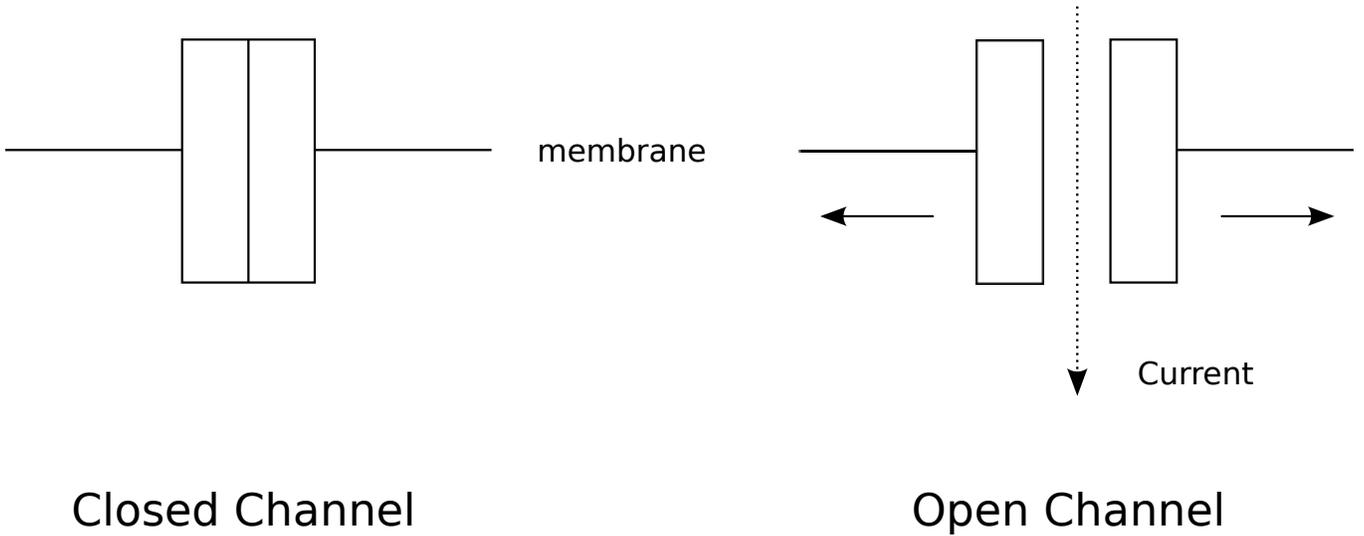}
\caption{Sketch of a stretch-activated channel in a closed and an open
state.}
\label{figChannel}
\end{figure}

A cartoon of a stretch activated ion channel is shown in Figure~\ref{figChannel}. The channel opens in response to deformation in the membrane.

The current through a population of stretch-activated ion channels may be modelled by
Ohm's law
\begin{gather}
I_M(\lambda,v)= g_M\big(\lambda(t)\big) \big(v-v_M\big),
\end{gather}
where $v_M$ is an equilibrium (aka ``reverse'') potential, and
$g_M\big(\lambda(t)\big)$ is a specific membrane conductance dependent
on the stretch $\lambda$. In a fully integrated model the stretch would be derived from nonlinear elasticity theory as detailed later in this report.

\subsection{Illustrative example}

\vspace{1mm}
For illustrative purposes and as a proof of concept, we now use
a set of caricature models to build a simplified model describing how the firing rate of afferent nerves depends on stretch.

\subsubsection{Caricature neural action potential model}

We consider McKean's model of an excitable system \cite{McKean1970}, to which we add a stretch activated current

\begin{align}
\label{eq3}
\frac{d v}{dt} & = -v
+H(v-a)-y+g_M\big(\lambda(t)\big)\big(v-v_M\big), \\
\frac{d y}{dt} &= \epsilon (v-d-cy),\nonumber
\end{align}
where $a$, $d$, $c$, and $\epsilon$ are constants.

\subsubsection{Caricature stretch-dependent conductance}

For the ``caricature'' stretch dependent conductance in \eqref{eq3}, we assume the following form 
\begin{gather}
g_M\big(\lambda(t)\big) =
\begin{cases}
\displaystyle
\frac{g_m}{1+k  \exp\big(-(\lambda-\lambda_c)\big)} \
&\text{if\ }
t< t_\text{empty},\\
0  \quad &\text{if\ } t \ge t_\text{empty},
\end{cases}\\
\end{gather}

where $g_m$ is the maximal conductance when the stretch $\lambda$ reaches is maximum $\lambda_c$, and $k$ modulates the dependence of $g_M$ on $\lambda$. We chose $\lambda$ to be related to the bladder radius, with $V_0$ the initial bladder volume and $Q_\text{in}$ the rate of bladder filling

\begin{equation}
\lambda=\left(1+\frac{Q_\text{in}}{V_0} t\right)^\frac{1}{3}.
\end{equation}

A plot of $g_M$ agaist time for a constant $Q_\text{in}$ is shown in Figure~\ref{figConductance}, where at $t=1800$ the bladder empties. 

\subsubsection{Parameter values}
The parameter values used in these simulations are given below. Please note that these parameters do not always have a physical meaning, and at this stage we have not allocated units or dimensions. The values have been adjusted to illustrate that model behaviour is broadly and qualitatively consistent with our knowledge of the physiology.
\begin{align*}
&
g_m=-0.7, \quad
k=1e2, \quad
\lambda_c=2, \quad
V_0=1,\quad
Q_\text{in}=1, \\
&
a= 0.3, \quad
\epsilon = 0.08,\quad
c = -0.3,\quad
d=-1/5,\quad
v_m = 1/2.
\end{align*}

\subsection{The Firing Neuron}

\begin{figure}
\includegraphics[width=0.6\textwidth]{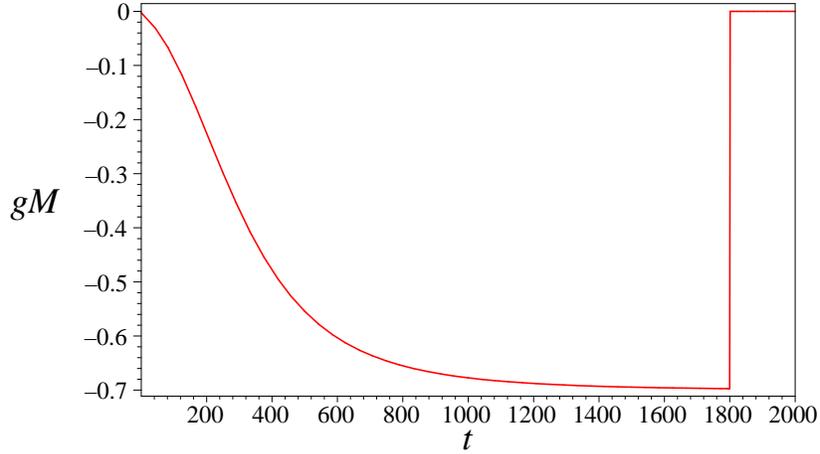}
\caption{The stretch-dependent conductance $g_M$ as a function of time.}
\label{figConductance}
\vspace{5mm}
\end{figure}

\begin{figure}
\hspace*{7mm}
\includegraphics[width=0.53\textwidth]{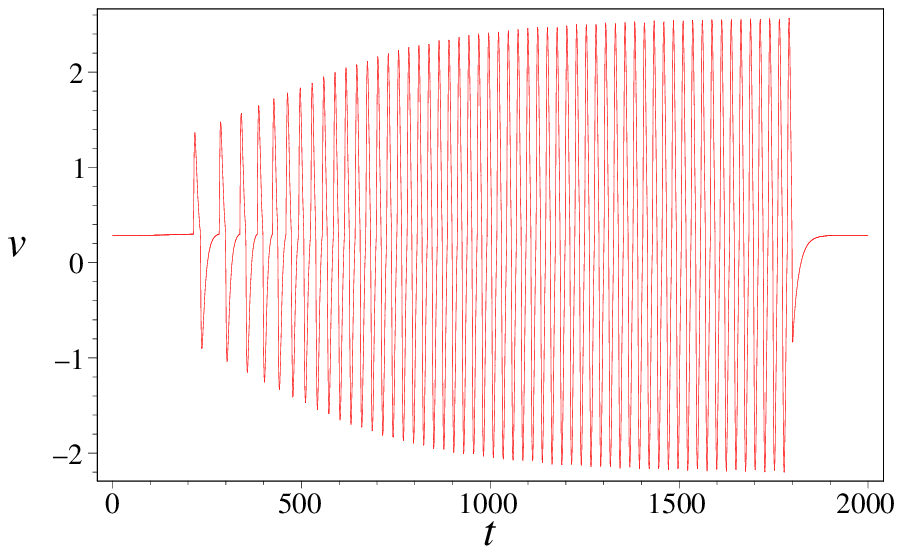}
\caption{Action potentials of a firing neuron in oscillatory state from \eqref{eq3}.}
\label{figAPs}
\end{figure}

An animation of the phase portrait of the McKean model can be viewed at \url{http://www.maths.gla.ac.uk/~rs/res/plot-PhasePortraitDeformation.gif}, which shows how the phase portrait changes as $g_M$ increases with time.

\begin{figure}
\hspace*{7mm}
\includegraphics[width=0.5\textwidth]{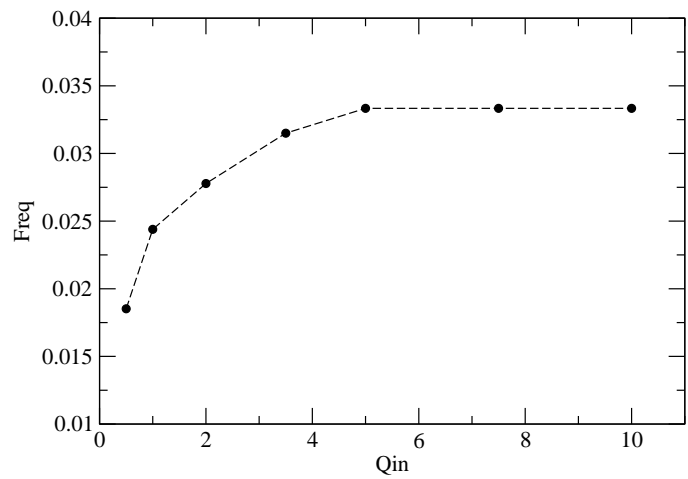}
\caption{Neuronal firing rate as a function of inflow rate computed
  from the model \eqref{eq3}.}
\label{figFiringRate}
\end{figure}

A time series of the model output is shown in Figure~\ref{figAPs}, which is plotted on the same timescale as Figure~\ref{figConductance}. This behaviour is qualitatively consistent with experimental data from the mouse bladder, and an example of these data is shown in Figure~\ref{NerveFiring1}. Experimental data also indicate that the rate of nerve firing also depends on the filling rate $Q_\text{in}$, as shown in Figure~\ref{NerveFiring2}. Our simple model also accounts for these changes, and Figure~\ref{figFiringRate} shows the sumpated dependence of firing rate on the bladder filling rate.

\newpage

\section{A quasi-linear viscoelastic model of the bladder wall}


\subsection{The problem}

An experiment was presented in which a bladder was filled and then held at a fixed volume whilst the pressure and afferent nerve activity within the bladder were measured as a function of time (see Figure~8, below). As can be observed in the figure, the pressure rapidly increased as the bladder was filled before slowly reducing once the filling had stopped. It can also be observed that the nerve activity appears to closely match the pressure, which suggests that at least some of the nerves are sensing the stress in the bladder wall as opposed to solely firing in response to strain. 

The phenomenon of stress reducing as a function of time given a fixed deformation is known as stress relaxation and is common to many viscoelastic materials. Linear viscoelasticity theory is commonly used to model the viscoelastic behaviour of materials under small strains, however, the bladder wall clearly undergoes a large deformation in the process of being inflated. Therefore, in the following section we model the bladder as a quasi-linear viscoelastic material, as described by \cite{DePascalis2014}. This theory is valid for materials undergoing large strains and is, therefore, more suited to this problem than the fully linear theory.

\begin{figure}
\includegraphics[width=0.9\textwidth]{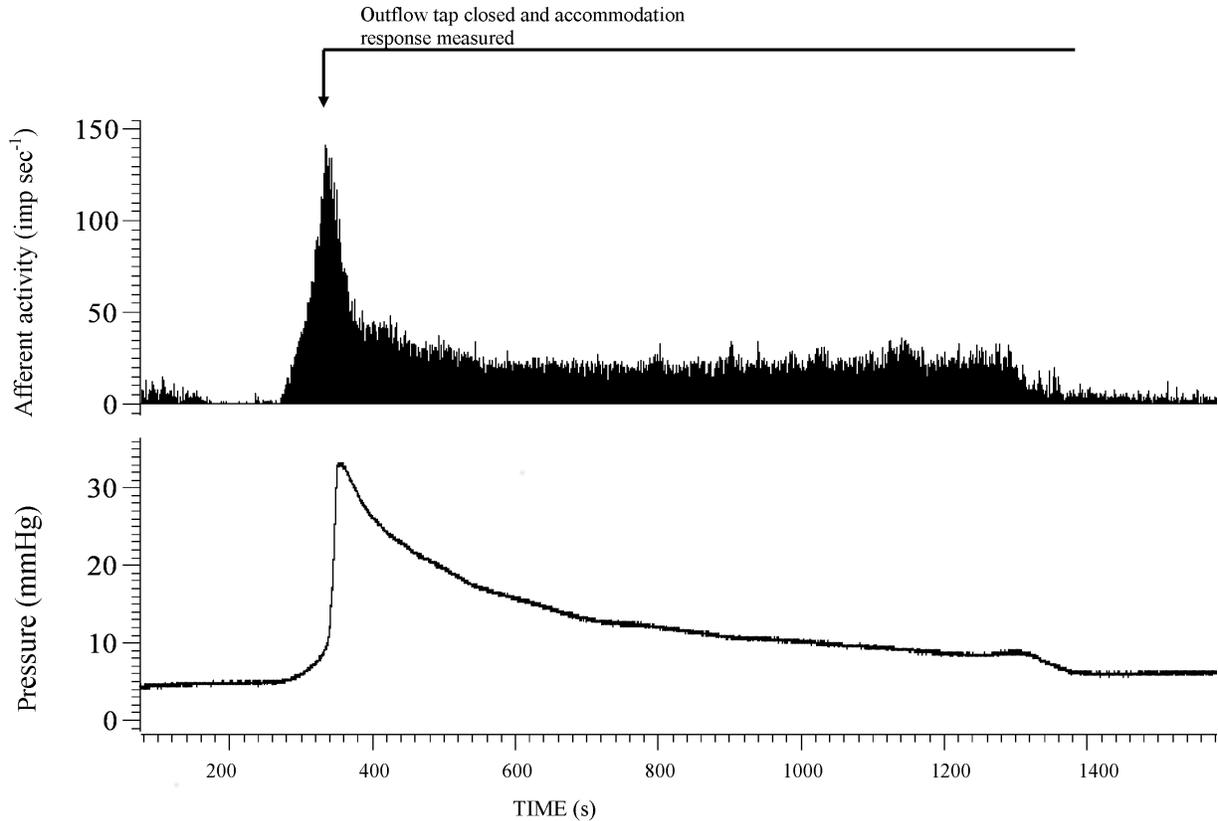}
\label{FigPressExpt}
\caption{A bladder was filled and then held at a constant volume. The resulting afferent nerve activity (top) and bladder pressure (bottom) are plotted as a function of time.}
\end{figure}

\subsection{Mathematical formulation}

We will model the bladder as an \textit{incompressible}, spherical shell with initial inner radius $A$ and outer radius $B$. We assume that the shell is isotropic and that its constitutive behaviour can be described by the reappraisal of Fung's quasi-linear viscoelastic model detailed in De Pascalis et al. (2014). We suppose that radial pressures are applied on the inner and outer surfaces of the shell, and that under such loading, the inner and outer  radii are deformed to $a$ and $b$, respectively. This deformation can be described by
\begin{equation}
 r=r(R,t),~~~\theta=\Theta,~~~\phi=\Phi,
\end{equation}
where $(R,\Theta,\Phi)$ and $(r,\theta,\phi)$ are spherical polar coordinates in the undeformed and deformed configurations respectively, and $r(R,t)$ is a function to be determined from the incompressibility condition. Position vectors in the undeformed and deformed conifgurations are
\begin{equation}
 \mathbf{X}=R\mathbf{E}_R(\Theta,\Phi),~~~\mathbf{x}=r(R,t)\mathbf{e}_r(\theta,\phi),
\end{equation}
where $\mathbf{E}_R$ and $\mathbf{e}_r$ are radial basis vectors in the undeformed and deformed configurations, respectively. The principal stretches associated with this deformation in the radial, azimuthal and polar directions are respectively
\begin{equation}
 \lambda_r=\frac{\partial r}{\partial R}~~~\lambda_\theta=\lambda_\phi=\frac{r}{R}.
\end{equation}
The deformation gradient tensor is defined by $\mathbf{F}=\Grad\mathbf{x}$, where $\Grad$ is the gradient operator associated with the undeformed configuration. In our case, we have
\begin{equation}
 \mathbf{F}(t)=F_{iJ}(t)\mathbf{e}_i\otimes\mathbf{E}_J,~~~F_{iJ}=\left(\begin{array}{ccc}
                                                                   \frac{\partial r}{\partial R} & 0 & 0\\
                                                                   0 & \frac{r}{R} & 0\\
                                                                   0 & 0 & \frac{r}{R}
                                                                  \end{array}\right),
\end{equation}
where $\mathbf{e}_i$, $i=(r,\theta,\phi)$, and $\mathbf{E}_J$, $J=(R,\Theta,\Phi)$ are the deformed and undeformed unit vectors in the radial, azimuthal and polar directions, respectively. For an incompressible material, we must have $J=\det\mathbf{F}=1$, and so
\begin{equation}
 \frac{r^2}{R^2}\frac{\partial r}{\partial R}=1.
\end{equation}
Solving the above, we obtain
\begin{equation}
 r(R,t)=(R^3+\alpha(t))^\frac{1}{3},
\end{equation}
where $\alpha$ is a constant defined by
\begin{equation}
 \alpha(t)=a^3(t)-A^3=b^3(t)-B^3.
\end{equation}
From (De Pascalis, et al., 2014), the time-dependent Cauchy stress tensor for an incompresible quasi-linear viscoelastic material is given by 
\begin{equation}
 \mathbf{T}(t)=\mathbf{F}(t)\left(\mathbf{\Pi}_D^e(t)+\int_0^t\mathcal{D}^\prime(t-s)\mathbf{\Pi}_D^e(s)ds\right)\mathbf{F}^\text{T}(t)-p(R,t)\mathbf{I},
\label{CauchyStress}
 \end{equation}
where $\mathbf{I}$ is the identity tensor, $\mathcal{D}(t)$ is a relaxation function and prime denotes differentiation with respect to the argument, $p(R,t)$ is a Lagrange multiplier associated with the incompressibility constraint which we have assumed to be independent of $\Theta$ and $\Phi$, and $\mathbf{\Pi}_D^e$ is the deviatoric component of the elastic second Piola-Kirchhoff stress tensor, which is defined by 
\begin{equation}
 \mathbf{\Pi}_D^e(t)=2\left(\frac{1}{3}(I_2W_2-I_1W_1)\mathbf{C}^{-1}(t)+W_1\mathbf{I}-W_2\mathbf{C}^{-2}(t)\right),
\label{PKstress}
\end{equation}
where $I_1$ and $I_2$ are strain invariants defined by
\begin{equation}
 I_1=\text{tr}\mathbf{B},~~~I_2=\frac{1}{2}((\text{tr}\mathbf{B})^2-\text{tr}\mathbf{B}^2),
\end{equation}
where $\mathbf{B}=\mathbf{FF}^\text{T}$ is the left Cauchy-Green tensor, $W_i=\partial W/\partial I_i$, where $W=W(I_1,I_2)$ is a strain energy function associated with the material's elastic response, and $\mathbf{C}=\mathbf{F}^\text{T}\mathbf{F}$ is the right Cauchy-Green stress tensor.

In order to consider the simplest possible case, we will assume that the material's \textit{elastic} response can be characterised by a neo-Hookean strain energy function:
\begin{equation}
 W=\frac{\mu}{2}(I_1-3).
\end{equation}
In this case, equation \eqref{PKstress} reduces to
\begin{equation}
 \mathbf{\Pi}_D^e(t)=\mu\left(\mathbf{I}-\frac{I_1}{3}\mathbf{C}^{-1}\right),
\end{equation}
where, in our case, $I_1$ is given by
\begin{equation}
 I_1=\frac{R^4}{(R^3+\alpha(t))^\frac{4}{3}}+2\frac{(R^3+\alpha(t))^\frac{2}{3}}{R^2},
\end{equation}
and $\mathbf{C}^{-1}$ is given by
\begin{equation}
 \mathbf{C}^{-1}=C^{-1}_{IJ}\mathbf{E}_I\otimes\mathbf{E}_J,~~~C_{IJ}^{-1}=\left(\begin{array}{ccc}
                                                                        \frac{(R^3+\alpha(t))^\frac{4}{3}}{R^4} & 0 & 0 \\
                                                                        0 & \frac{R^2}{(R^3+\alpha(t))^\frac{2}{3}} & \\
                                                                        0 & 0 & \frac{R^2}{(R^3+\alpha(t))^\frac{2}{3}}  
                                                                       \end{array}\right).
\end{equation}
Using the above in \eqref{CauchyStress}, we can derive explicit expressions for the non-zero Cauchy stresses (where $\mathbf{T}=T_{ij}\mathbf{e}_i\otimes\mathbf{e}_j$):
\begin{multline}
 T_{rr}(t)=\mu\left(\frac{R^4}{(R^3+\alpha(t))^\frac{4}{3}}-\frac{I_1}{3}\right)+\\
 \frac{\mu R^4}{(R^3+\alpha(t))^\frac{4}{3}}\int_0^t\mathcal{D}^\prime(t-s)\left(1-\frac{I_1}{3}\frac{(R^3+\alpha(s))^\frac{4}{3}}{R^4}\right)ds-p(R,t),
\end{multline}
\begin{multline}
 T_{\theta\theta}(t)=T_{\phi\phi}(t)=\mu\left(\frac{(R^3+\alpha(t))^\frac{4}{3}}{R^4}-\frac{I_1}{3}\right)+\\
 \frac{\mu(R^3+\alpha(t))^\frac{4}{3}}{R^4}\int_0^t\mathcal{D}^\prime(t-s)\left(1-\frac{I_1}{3}\frac{R^4}{(R^3+\alpha(s))^\frac{4}{3}}\right)ds-p(R,t).
\end{multline}
If we neglect body forces, then the equations of equilibrium are given by
\begin{equation}
 \text{div}\mathbf{T}=\rho\frac{\partial^2\mathbf{U}}{\partial t^2},
\end{equation}
where $\text{div}$ is the divergence operator in the deformed configuration, $\rho$ is the density of the material under consideration, and $\mathbf{U}(t)$ is the displacement vector, defined by $\mathbf{U}=\mathbf{x}-\mathbf{X}$. For our deformation, the only equation not trivially satisfied is the radial equation:
\begin{equation}
 r\frac{d}{dr}(T_{rr})+2(T_{rr}-T_{\theta\theta})=\rho\frac{\partial^2U}{\partial t^2},
\end{equation}
where $U(t)=r(t)-R=(R^3+\alpha(t))^\frac{1}{3}-R$, and we have used the fact that $T_{\theta\theta}=T_{\phi\phi}$. By rearranging the above equation, and applying the boundary conditions $T_{rr}|_{r=a}=-p(t)$, $T_{rr}|_{r=b}=-p_e$ (we have assumed that $p_e$ is fixed for all time), we may obtain a relationship between the pressure difference, $p_i(t)-p_e$, and the deformation parameter $\alpha(t)$ as a function of time:
\begin{multline}
 p(t)-p_e=\int_A^B\left(\rho\frac{R^2}{R^3+\alpha(t)}\frac{\partial^2U}{\partial t^2}+2\mu\left(\frac{(R^3+\alpha(t))^\frac{1}{3}}{R^2}-\frac{R^6}{(R^3+\alpha(t))^\frac{7}{3}}+\right.\right.\\
 \left.\left.\frac{(R^3+\alpha(t))^\frac{1}{3}}{R^2}\int_0^t\mathcal{D}^\prime(t-s)\left(1-\frac{I_1}{3}\frac{R^4}{(R^3+\alpha(s)))^\frac{4}{3}}\right)ds-\right.\right.\\
 \left.\left.\frac{R^6}{(R^3+\alpha(t))^\frac{7}{3}}\int_0^t\mathcal{D}^\prime(t-s)\left(1-\frac{I_1}{3}\frac{(R^3+\alpha(s))^\frac{4}{3}}{R^4}\right)ds\right)\right)dR.
 \label{PressDefEqn}
\end{multline}
To proceed, we must select a relaxation function $\mathcal{D}$. In this case, as in (De Pascalis et al., 2014), we shall use the one term Prony series:
\begin{equation}
 \mathcal{D}(t)=\frac{\mu_\infty}{\mu}+\left(1-\frac{\mu_\infty}{\mu}\right)e^{-\frac{t}{\tau}},
 \label{PronySeries}
\end{equation}
where $\mu_\infty$ is the long-time infinitesimal shear modulus, and $\tau$ is the relaxation time.

Upon substituting equation \eqref{PronySeries} into equation \eqref{PressDefEqn}, we obtain an explicit relationship between the applied deformation and the pressure difference required to maintain it. In the following section we prescribe the radial deformation parameter $\alpha(t)$, which we note has dimension $[\text{length}^3]$, and plot the resulting inner radial pressure as a function of time.

\subsection{Results}

In this section we show that the model described above qualitatively exhibits the same behaviour as the experimental data shown in Figure~8. Since independent measurements of the parameters required for the model were not available, these were chosen in order to fit the data. The assumed form for $\alpha(t)$ was
\begin{equation}
\alpha(t)=182\tanh(0.03(t-320\text{s}))\text{ml},
\end{equation}
and is plotted in Figure~9, below. The initial inner and outer radii were chosen to be $A=4\text{cm}$ and $B=5\text{cm}$. Given these parameters, the resulting bladder volume can be derived:
\begin{equation}
V(t)=\frac{4\pi}{3}a^3(t)=\frac{4\pi}{3}(A^3+\alpha(t)).
\end{equation}
In Figure~10, we plot the bladder volume as a function of time.

\begin{figure}
\includegraphics[width=0.6\textwidth]{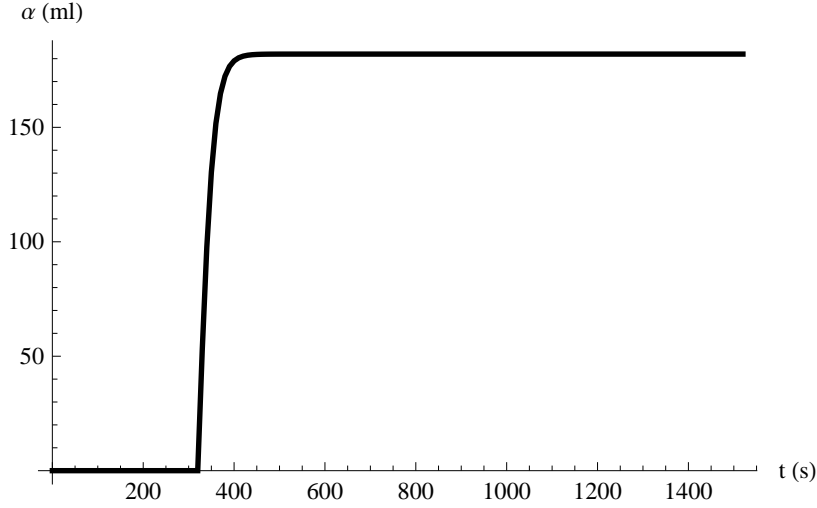}
\label{figalpha}
\caption{The deformation parameter $\alpha(t)$ as a function of time.}
\end{figure}

\begin{figure}
\includegraphics[width=0.6\textwidth]{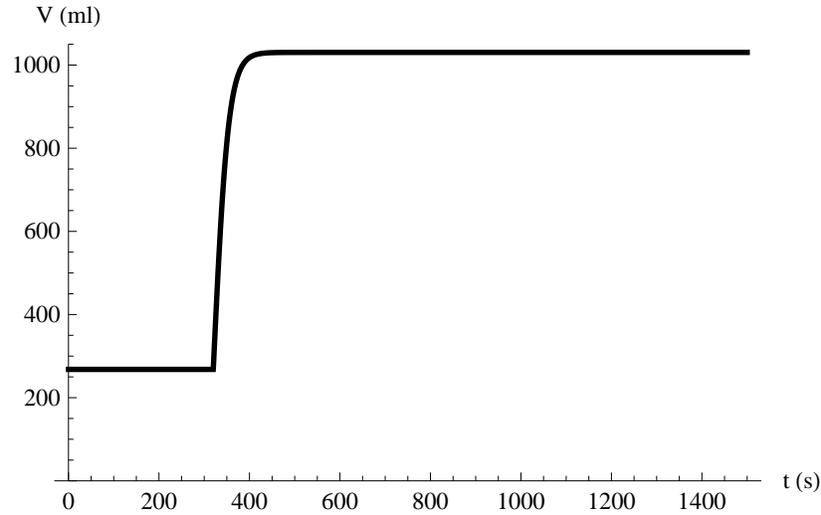}
\label{figvolume}
\caption{The bladder volume $V(t)$ as a function of time.}
\end{figure}

Finally, we must select values for the remaining parameters in order to plot the inner pressure $p(t)$ as a function of time. In this case, we have chosen $\mu=48\text{mmHg}\approx 6.4\text{kPa}$, $\mu_\infty=1\text{mmHg}\approx 0.1\text{kPa}$, $\rho=1\text{kg}/\text{m}^3$, $\tau=5$ and $p_e=5\text{mmHg}\approx 0.7\text{kPa}$. Given these parameter values, we may numerically evaluate the integrals in equation \eqref{PressDefEqn} in order to determine $p(t)$. The numerical solver used in this case was \texttt{NIntegrate} in \textit{Mathematica 7}. The results are plotted against the experimental data in Figure~11. As can be seen, the theoretical results show qualitative agreement with the experimental data, which suggests that the viscoelasticity of the bladder wall may be responsible for the observed pressure drop.

\begin{figure}
\includegraphics[width=0.6\textwidth]{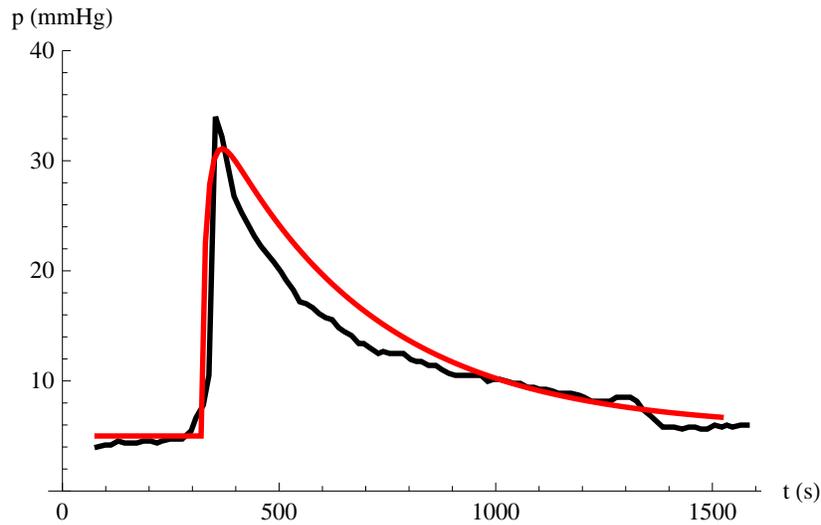}
\label{figresults}
\caption{The inner pressure $p(t)$ as a function of time. Theoretical results: red, experimental data: black.}
\end{figure}

\subsection{Future work}

We note that the model derived above contains a combination of measurable and phenomenological parameters, as listed in the table below:

\begin{center}
\begin{tabular}{|c|c|}
\hline
Measurable parameters & Phenomenological parameters\\
\hline
$\alpha(t)$ - radial deformation parameter & $\mu$ - instantaneous shear modulus\\
$A$ - initial inner bladder radius & $\mu_\infty$ - long-time shear modulus\\
$B$ - initial outer bladder radius & $\tau$ - relaxation time\\
$\rho$ - bladder wall density & \\
$p_e$ - external pressure & \\
\hline
\end{tabular}
\end{center}

In order to fully test the model, it will be important to independently calculate all of the measurable parameters. In the long term, arguments based on the microstructure of the various layers of the bladder wall could be used to replace the phenomenological parameters $\mu$, $\mu_\infty$ and $\tau$ with parameters that can be directly measured.

\section{A finite element model of filling and emptying of bladder }


A 2D mathematical model was developed on the basis of the problem described above in Section 5.1 using Lame's system of elasticity.

\subsection{Mathematical formulation}

Solid objects deform under the action of applied forces in such a way that a point in the solid, originally at $(x,y)$ would come to $(X,Y)$ after some time, thus, resulting in displacement in the form of vector ${\bf u}=(u_1, u_2)=(X-x, Y-y)$. Hooke’s law gives a relationship between the stress tensor $\sigma({\bf u}) =(\sigma_{ij}(u))$ and the strain tensor $\epsilon({\bf u})=\epsilon_{ij}({\bf u})$.
\begin{eqnarray}
\sigma_{ij} &=& \lambda \delta_{ij} \nabla \cdot {\bf u} + 2\mu \epsilon_{ij}({\bf u})
\end{eqnarray}
where the Kronecker symbol $\delta_{ij}=1$ if $i=j$, $0$  otherwise, with
\begin{eqnarray}
\epsilon_{ij} &=& \frac{1}{2}\left(\frac{\partial u_i}{\partial x_j}+ \frac{\partial u_j}{\partial x_i}\right)
\end{eqnarray}
and where $\lambda$, $\mu$ are the Lame's constants that describe $E$, the Young’s modulus, and  $\nu$, the Poisson’s ratio via
\begin{eqnarray}
\mu &=& \frac{E}{2\left(1+\nu\right)},\\
\lambda &=& \frac{E \nu}{\left(1+\nu\right)\left(1-2\nu\right)}.
\end{eqnarray}

In a 2D domain, the components along $x$ and $y$ of the strain $u$ in section $\Omega$ subjected to force $f$ perpendicular to the axis is governed by:
\begin{eqnarray}
-\mu \nabla^2 {\bf u}-\left(\mu+\lambda\right) \nabla\left(\nabla \cdot \bf{u}\right) = f \mbox{ in } \Omega,\label{eqn41}
\end{eqnarray}

We do not use equation (\ref{eqn41}) because the associated variational or  weak form does not give the right boundary conditions, therefore, we use:
\begin{eqnarray}
-\nabla \cdot \sigma = f \mbox{ in } \Omega.\label{eqn42}
\end{eqnarray}

\paragraph{Definition of Weak or Variational formulation }
It is an important tool for the analysis of mathematical equations that permit the transfer of concepts of linear algebra to solve problems in other fields such as partial differential equations. In a weak formulation, an equation is no longer required to hold absolutely (and this is not even well defined) and has instead weak solutions only with respect to certain ``test vectors'' or ``test functions''. This is equivalent to formulating the problem to require a solution in the sense of a distribution.

The weak formulation for (\ref{eqn42}) is given by
\begin{eqnarray}
\int_{\Omega} \sigma({\bf u}):\epsilon ({\bf v}) d{\bf x}-\int_{\Omega} {\bf v} f d{\bf x}=0, 
\end{eqnarray}
where $:$ denotes the tensor scalar product, i.e. $a:b=\Sigma_{ij} a_{ij} b_{ij}$ and ${\bf v}$ is a test function.  The variational form can then be written as
\begin{eqnarray}
\int_{\Omega} \lambda \nabla \cdot {\bf u} \nabla \cdot {\bf v} + 2 \mu \epsilon ({\bf u}): \epsilon ({\bf v}) d{\bf x} - \int_{\Omega} {\bf v} f d{\bf x} = 0.
\end{eqnarray}

\subsection{Initial and Boundary conditions}

\begin{figure}
\includegraphics[width=0.7\textwidth]{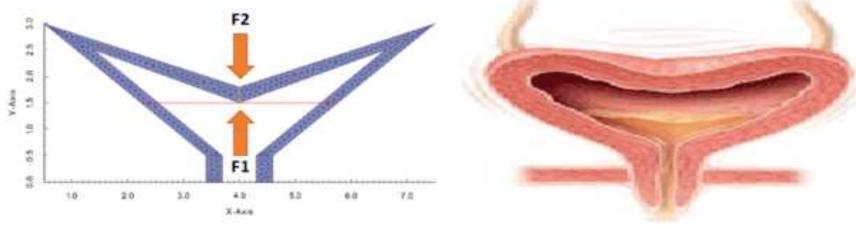}
\caption{(Left) 2D finite element (FE) domain of the bladder wall. (Right) Schematic diagram of the bladder wall.}\label{figVarun1}
\end{figure}

Let us consider a 2D cross section (7mm x 3mm) of a bladder in the transverse plane (see fig. \ref{figVarun1}) with bladder wall thickness of 0.3 mm. The Young’s modulus and the Poisson’s ration of the bladder was taken to be as 10 kPa and 0.49 respectively (Hensel et al., 2007).  The bladder wall is subjected to forces $F_1$ (dominant force during filling phase) and $F_2$ (dominant force during emptying phase) perpendicular to the x-axis. The bottom of the domain is fixed.

\subsection{Results}
In this section, we show that the filling and emptying phase of bladder can be modelled using the finite element method. The model was developed and implemented in FreeFEM++, a free PDE solver based on C++. 

\begin{figure}
\includegraphics[width=0.7\textwidth]{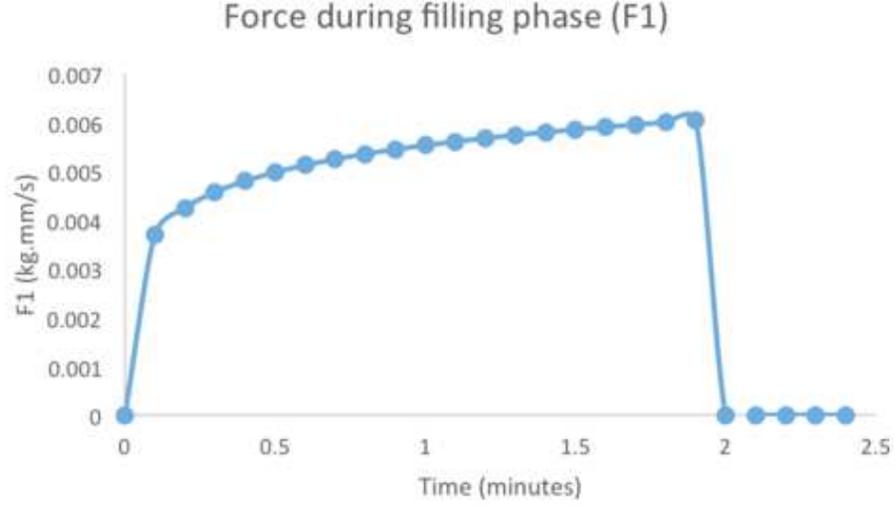}
\caption{$F_1$ as a function of time.}\label{fig10}
\end{figure}

\begin{figure}
\includegraphics[width=0.7\textwidth]{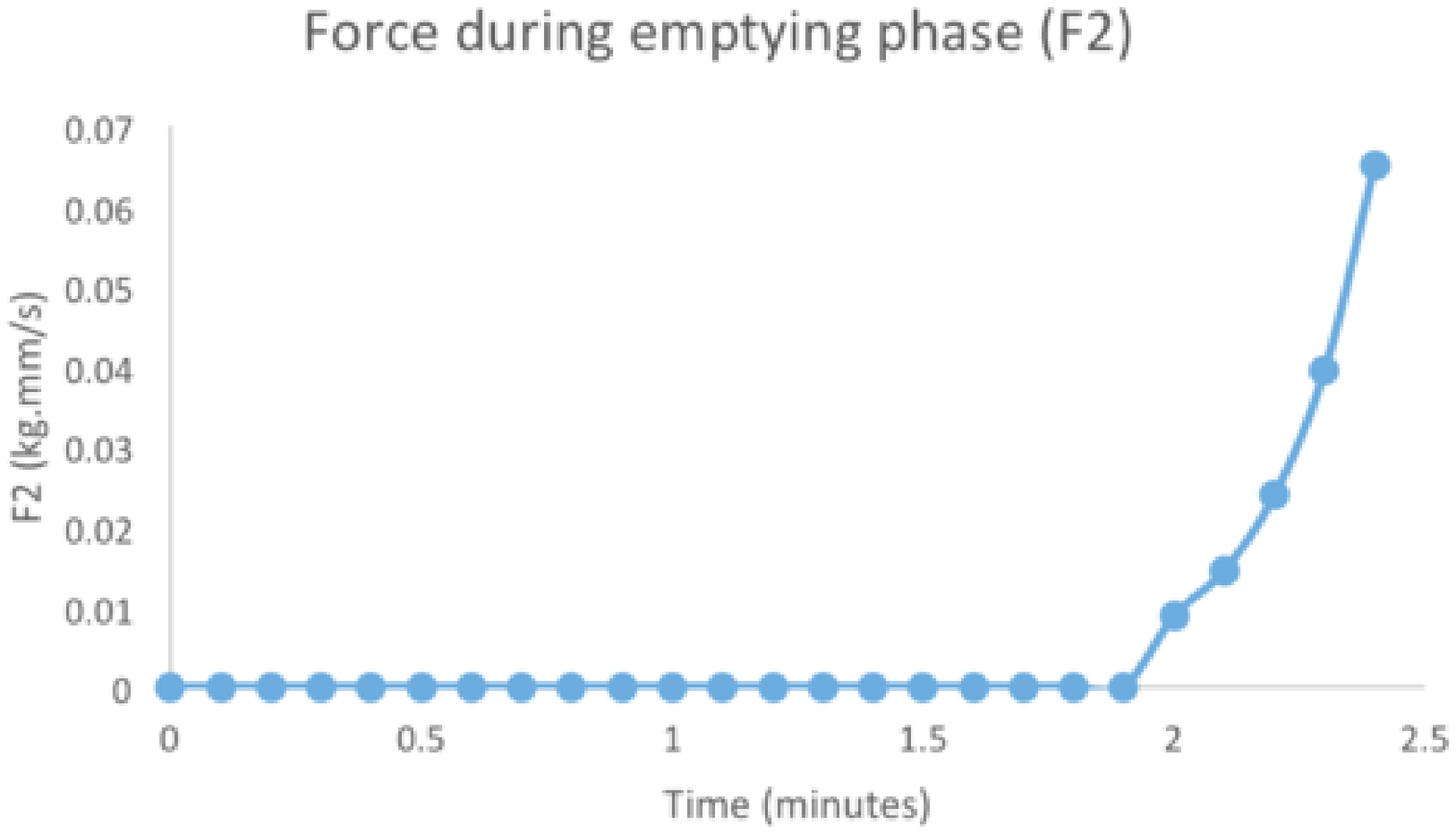}
\caption{$F_2$ as a function of time.}\label{fig11}
\end{figure}

\begin{figure}
\includegraphics[width=0.7\textwidth]{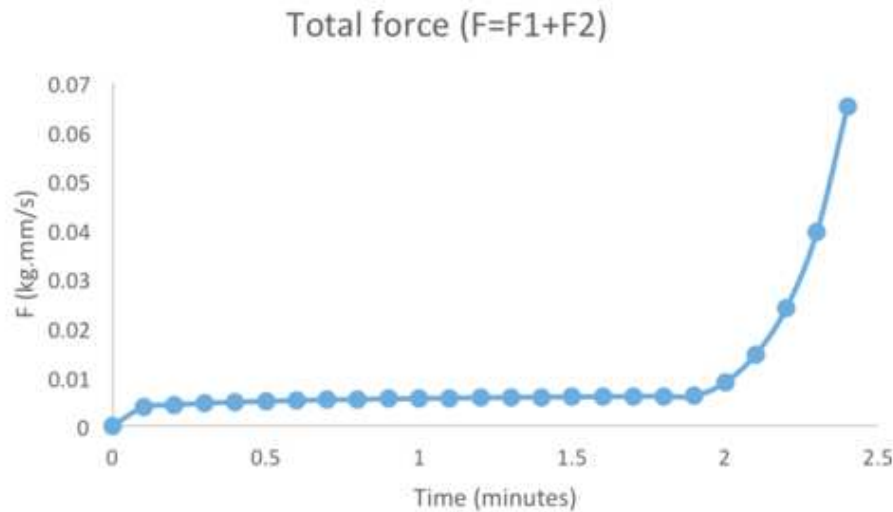}
\caption{Total force acting on the bladder wall as a function of time.}\label{fig12}
\end{figure}

\begin{figure}
\includegraphics[width=0.7\textwidth]{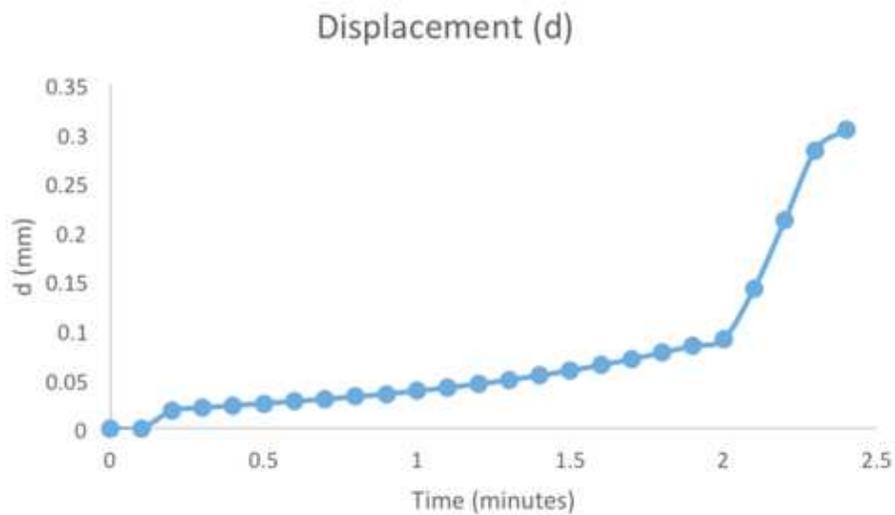}
\caption{Displacement as a function of time.}\label{fig13}
\end{figure}

\begin{figure}
\includegraphics[width=\textwidth]{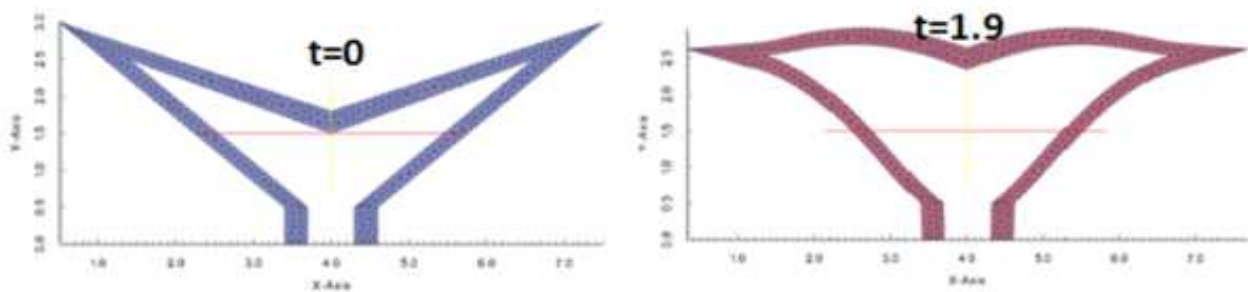}
\caption{Position of bladder wall at time t=0 and t=1.9 minutes.}\label{fig14}
\end{figure}

\begin{figure}
\includegraphics[width=0.7\textwidth]{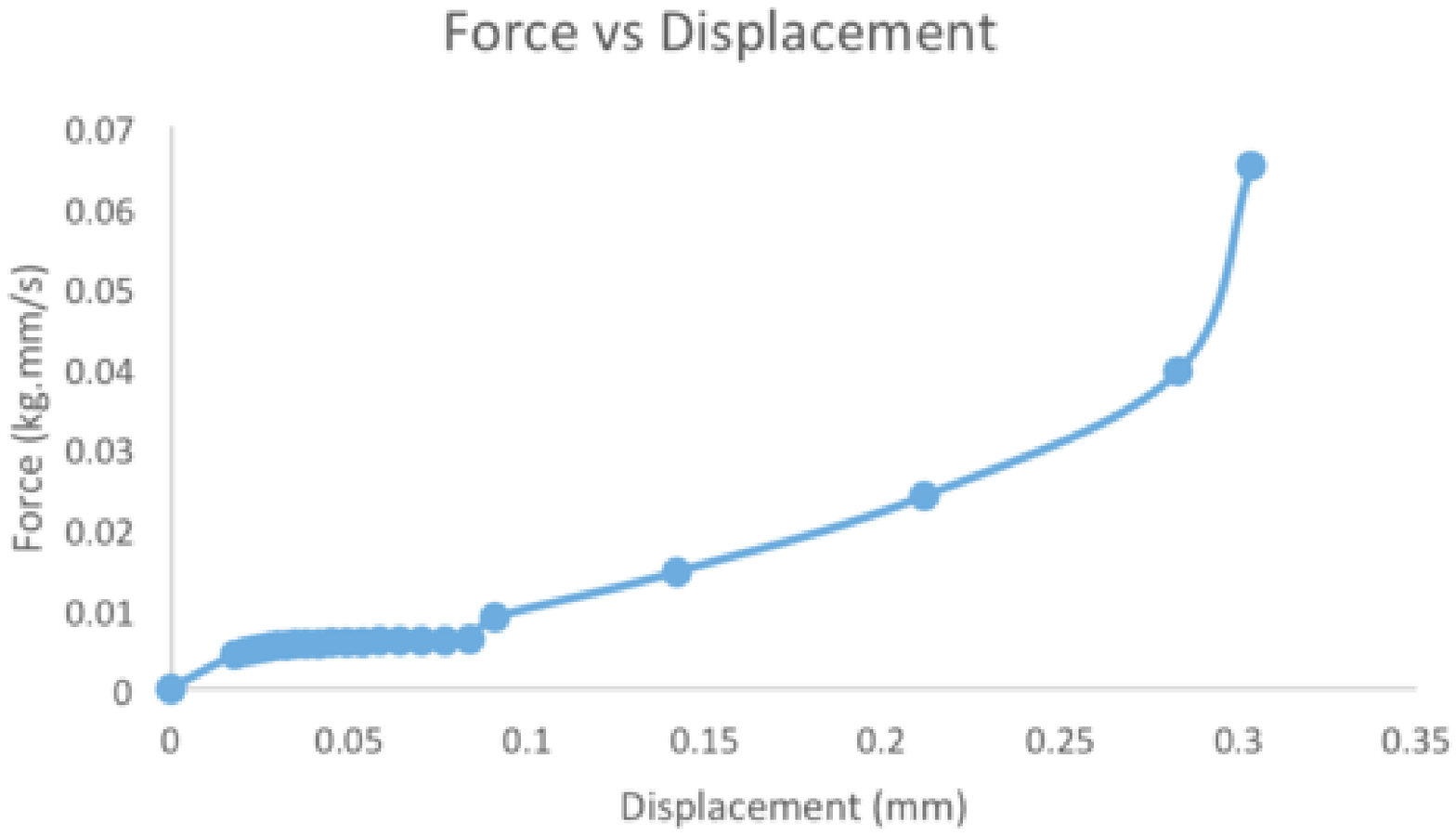}
\caption{Force versus displacement curve.}\label{fig15}
\end{figure}

As the forces acting on the bladder wall during the filling phase and the emptying phase were not available, the forces mentioned below were chosen to fit the data. The forces (see fig. \ref{fig10} and fig. \ref{fig11}) were made dependent on displacement ($d$) of the bladder wall and set as function of time such that
\begin{eqnarray}
F_1 & = & 0.0008 \log (1+1000 t), \mbox{ if } d< 0.09 mm\\
F_2 & = & 0.0000004 \exp(5t), \mbox{ if } d> 0.09 mm. 	
\end{eqnarray}
Therefore, the total force (see fig. \ref{fig12}) acting on the bladder during filling and emptying process can be calculated as total force $F=F_1+F_2$. The force acting on the wall leads to displacement ($d$) which has been plotted below over time (see fig. \ref{fig13}). In fig. \ref{fig14} we see the maximum displacement in the bladder wall during the filling process.  Finally, we could plot the relationship between the force and the deformation during bladder filling and emptying process (see fig. \ref{fig15}).

\subsection{Future work}
In order to fully test the model, a smoothed shape of the bladder wall has to be considered for the finite element analysis. The model has to consider bladder wall to be viscoelastic in nature and not just elastic as in the presented model. Also, the magnitude of forces acting on the bladder wall have to be carefully measured to validate the assumptions made in the model. The forces acting on the bladder wall could also be made dependent on the nerve response. Finally, the model could be coupled with mechanics of the fluid which would involve solving fluid-structure interactions (FSI).

\section{Towards an integrative model}

During the study group week, several models describing the mechanics of the bladder wall and the influence of stretch on nerve activity were developed, and have been described above. While each of these models could be developed further in order to investigate specific aspects of bladder mechanics, the most exciting prospect is to combine the models so that an integrative model that describes the filling fo the bladder could be developed. Other models of bladder function have been described in the literature \cite{Bastianssen1996, Valentini2000}, but the focus of these models has been on the emptying phase rather than bladder filling. Some studies of bladder filling include simplified models of viscoelasticity as well as experimental data on model parameters \cite{vanMastrigt1978, vanMastrigt1981}, which could form the basis for extending and parameterising the models developed during the study group.

A specific next stage towards integration would be to reconstruct the experimental mouse bladder model that was the starting point of this study group problem \cite{Daly2014}. This could probably be achieved without any additional experiments.

\section{Discussion and conclusions}

The application of mathematics and engineering to describe and understand the structure and function of the human bladder is an area with enormous potential. During the study group several preliminary models were developed, which were able to qualitatively reporoduce experimental findings, demonstraing that models of stretch sensitive channels and viscoelasticity could account for experimental observations.

\subsection{Potential NC3Rs benefits}

Animals have been used in urology research since the 1930s to 1) model human diseases; 2) study normal physiological processes and 3) provide a source of biological material such as tissues and cells for scientific analysis. In the first two of these objectives, rodents have been extremely valuable as they allow an integrative system approach to studying normal and pathological function. Unfortunately, replacement of animals entirely in this regard is still not a possibility; however, reduction of animal numbers and methodological refinement are being pursued by the scientific community. Currently, studying bladder afferent nerve pathways relies exclusively on the use of in vitro or in vivo animal models (usually rodents). There are no official estimates on the numbers of animals used for these experiments however there are around 800 papers per year in the field of urology (excluding reviews and clinical trials). A Pubmed search with the keywords “bladder and afferent” but not “human” produced 161 papers from 2008-2013 (all figures exclude reviews). Assuming that each study used about 40-50 animals then the estimated number of animals sacrificed each year for bladder research could exceed 1300. If we were to consider animals used in unpublished academic and pharmaceutical research then the actual numbers used are probably several times higher. More alarmingly, this number could be several times higher if we were to consider the animals used for unpublished academic and pharmaceutical research. Moreover, OAB and UI have increased prevalence in the elderly. As western populations are demographically ageing, there is likely to be considerable growth in the research field over the next couple of years, especially now that ageing research has been prioritised by the research councils.

Developing a mathematical model to study sensory nerve pathways would significantly reduce the need for animal use and may even in some cases replace the use of in vivo/in vitro models all together.

\subsection{Potential healthcare benefits}

Functional disorders of the lower urinary tract place a huge burden on global healthcare resources. In the UK, treatment and management of OAB and UI costs the NHS in excess of £233 million per year. Although not life threatening, these conditions severely impair the sufferers’ quality of life, leading to sleep deprivation, depression, embarrassment and fatigue. The underlying aetiology of these disorders are unknown, but there is clear correlation between prevalence of OAB and UI, and age. Moreover incontinence has been cited as the major reason for institutionalisation of the elderly. Given that western populations are demographically ageing, there is an urgent need to understand the mechanisms that lead to these conditions and to identify new drug targets for treatment.

The current mainstay for OAB/UI treatment is the use of drugs which inhibit bladder contractility; however, these drugs are poorly tolerated, relatively ineffective and can cause bladder retention (i.e. compromised ability to empty the bladder). Moreover, the cardinal symptoms of OAB: urgency (urgent desire to urinate) and frequent urination, appear to be driven by changes in sensory nerve pathways and it has been suggested that the sensory nerves may be a target for future therapeutic interventions. Developing a mathematical model which can be used to look at these pathways will provide an entirely novel tool in the research armoury, facilitating and accelerating research progress and aiding in the screening and development of new pharmaceutical agents.

\end{document}